\documentclass[jkps,preprint,fleqn,showpacs,showkeys,superscriptaddress]{revtex4}
\usepackage[pdftex]{graphicx}
\usepackage{amssymb}
\usepackage{amsmath}
\usepackage{bm}
\usepackage{lineno}

\begin{document}
\setcounter{page}{0}
\title[]{Measurement of Fast Neutron Rate for NEOS Experiment}
\affiliation{Department of Physics, Chonnam National University, Gwangju 61186, Korea}
\affiliation{Department of Physics, Chung-Ang University, Seoul 06974, Korea}
\affiliation{Center for Underground Physics, Institute for Basic Science, Daejeon 34047, Korea}
\affiliation{Neutron Science Division, Korea Atomic Energy Research Institute, Daejeon 34057, Korea}
\affiliation{Korea Research Institute of Standards and Science, Daejeon 34113, Korea}
\affiliation{Department of Physics, Kyungpook National University, Daegu 41566, Korea}
\affiliation{Department of Physics and Astronomy, Sejong University, Seoul 05006, Korea}
\affiliation{University of Science and Technology, Daejeon 34113, Korea}

\author{Y.~J.~Ko}
\affiliation{Department of Physics, Chung-Ang University, Seoul 06974, Korea}

\author{J.~Y.~Kim}
\affiliation{Department of Physics and Astronomy, Sejong University, Seoul 05006, Korea}

\author{B.~Y.~Han}
\affiliation{Neutron Science Division, Korea Atomic Energy Research Institute, Daejeon 34057, Korea}

\author{C.~H.~Jang}
\affiliation{Department of Physics, Chung-Ang University, Seoul 06974, Korea}

\author{E.~J.~Jeon}
\affiliation{Center for Underground Physics, Institute for Basic Science, Daejeon 34047, Korea}

\author{K.~K.~Joo}
\affiliation{Department of Physics, Chonnam National University, Gwangju 61186, Korea}

\author{B.~R.~Kim}
\affiliation{Department of Physics, Chonnam National University, Gwangju 61186, Korea}

\author{H.~J.~Kim}
\affiliation{Department of Physics, Kyungpook National University, Daegu 41566, Korea}

\author{H.~S.~Kim}
\email{hyunsookim@sejong.ac.kr}
\thanks{Fax:+82-2-3408-4316}
\affiliation{Department of Physics and Astronomy, Sejong University, Seoul 05006, Korea}

\author{Y.~D.~Kim}
\affiliation{Center for Underground Physics, Institute for Basic Science, Daejeon 34047, Korea}
\affiliation{Department of Physics and Astronomy, Sejong University, Seoul 05006, Korea}
\affiliation{University of Science and Technology, Daejeon 34113, Korea}

\author{Jaison Lee}
\affiliation{Center for Underground Physics, Institute for Basic Science, Daejeon 34047, Korea}

\author{J.~Y.~Lee}
\affiliation{Department of Physics, Kyungpook National University, Daegu 41566, Korea}

\author{M.~H.~Lee}
\affiliation{Center for Underground Physics, Institute for Basic Science, Daejeon 34047, Korea}

\author{Y.~M.~Oh}
\affiliation{Center for Underground Physics, Institute for Basic Science, Daejeon 34047, Korea}

\author{H.~K.~Park}
\affiliation{Center for Underground Physics, Institute for Basic Science, Daejeon 34047, Korea}
\affiliation{University of Science and Technology, Daejeon 34113, Korea}

\author{H.~S.~Park}
\affiliation{Korea Research Institute of Standards and Science, Daejeon 34113, Korea}

\author{K.~S.~Park}
\affiliation{Center for Underground Physics, Institute for Basic Science, Daejeon 34047, Korea}

\author{K.~M.~Seo}
\affiliation{Department of Physics and Astronomy, Sejong University, Seoul 05006, Korea}

\author{Kim Siyeon}
\affiliation{Department of Physics, Chung-Ang University, Seoul 06974, Korea}

\author{G.~M.~Sun}
\affiliation{Neutron Science Division, Korea Atomic Energy Research Institute, Daejeon 34057, Korea}

\collaboration{NEOS Collaboration}
\noaffiliation

\date{\today}
\begin{abstract}
The fast neutron rate is measured at the site of NEOS experiment, a short baseline 
neutrino experiment located in a tendon gallery of a commercial nuclear power plant, 
using a 0.78-liter liquid scintillator detector. A pulse shape 
discrimination technique is used to identify neutron signals. The measurements are 
performed during the nuclear reactor-on and off periods and 
found to be $\sim20$ per day for both periods. 
The fast neutron rate is also measured at an overground site with a negligible 
overburden and is found to be $\sim 100$ times higher than that at the NEOS 
experiment site.
\end{abstract}
\pacs{68.37.Ef, 82.20.-w, 68.43.-h}
\keywords{fast neutron rate, neutron background, short baseline, reactor antineutrino, sterile neutrino}
\maketitle

\section{INTRODUCTION}
Short baseline reactor neutrino experiments such as NEOS (Neutrino Experiment 
Oscillation Short baseline) experiment are designed to search 
for the possible existence of sterile neutrinos~\cite{Bowden:2016ntq}. 
One of the motivations for these experiments 
is so called ``reactor anomaly''~\cite{Mention:2011rk} and it is observed by other reactor 
based neutrino experiments~\cite{An:2015nua,RENO:2015ksa}. 
The electron antineutrinos from the reactor can be detected by the inverse beta decay (IBD),
${\bar \nu}_e + p \rightarrow e^+ + n$. 
In a liquid scintillator (LS) doped with Gd, which is typically used for reactor based neutrino 
experiments, the positron produces light signal immediately giving the prompt signal and 
the neutron gets thermalized and then captured by either H or Gd. Subsequently, a delayed 
signal, a 2.2~MeV photon signal from H captured event or $\sim 8$~MeV from Gd captured event, 
follows the prompt signal by on the order of tens of microseconds.
The existence of the sterile neutrinos can be probed by examining the rate of IBD events.

To maximize the sensitivity to the existence of the sterile neutrinos in the reactor based 
neutrino experiments, the detectors need to be placed as close to the reactor as possible,
on the order of tens of meters or less.
However, it is often hard to find a space to install the detector 
at such a close range to the reactor 
with adequate shielding from the cosmogenic backgrounds as well as the reactor originated
neutrons to a lesser extent. 
The fast neutrons can mimic the IBD signal;
the neutron elastic scattering produces a recoiling proton that can be 
misidentified as a prompt signal and then captured by H or Gd to produce 
a delayed signal.
Therefore, finding a suitable location to place the detector with enough overburden and 
shielding is important.

Table~\ref{tb:shrotbaseline} lists the recent short baseline neutrino experiments.
DANSS and NEOS experiments use commercial nuclear power plant reactors, whereas
the remaining experiments use research reactors. 
Generally a commercial reactor based experiment is provided with a larger neutrino 
flux and a better shielding against neutrons due to heavy structures the reactor 
complex is required to have. 
While there is more space available for installing the detector at research reactor 
facilities for a better sensitivity, the research reactors have a small neutrino flux than 
that of the commercial reactor based experiment. 
Also, the research reactor
facilities usually have a negligible overburden and they are primarily used as a neutron source.   
In fact, the NUCIFER experiment has observed a significant difference in the neutron rate
between the reactor-on and off periods~\cite{Boireau:2015dda}.  
Therefore, it is important to measure the neutron rates at the possible experiment sites.
\begin{table}[ht]
\caption{Overview of reactor based short baseline neutrino experiments.}
\begin{ruledtabular}
\begin{tabular}{cccc}
Experiment & Reactor Thermal Power & Baseline & Location \\
\colrule
NUCIFER~\cite{Boireau:2015dda}          & 70~MW  & $\sim7$~m               & France \\
PROSEPCT~\cite{Ashenfelter:2015aaa} & 85~MW  & 7-10~m (near)/15-19~m (far) & US \\
STEREO~\cite{Zsoldos:2016bkr}       & 57~MW  & 9-11~m                      & France \\
CHANDLER~\cite{Bowden:2016ntq}      & 60~MW  & $\sim6$~m                   & Belgium \\
NEOS~\cite{Bowden:2016ntq}          & 2.8~GW & 24~m                        & Korea \\
DANSS~\cite{Alekseev:2016llm}       & 3.0~GW   &11-13~m                    & Russia \\
\end{tabular}
\end{ruledtabular}
\label{tb:shrotbaseline}
\end{table}

We present measurements of the fast neutron rate using a pulse shape discrimination (PSD) 
technique at the NEOS experiment site, which is
located in a tendon gallery of Hanbit nuclear power plant complex, and an overground site 
with a negligible overburden to emulate the measurements at research reactor sites. 
The measurements were performed before the site selection. 
Based on many criteria, especially on the fast neutron rate among others, a site at Hanbit nuclear 
power plant was chosen over a research reactor facility. 
The NEOS experiment took data for an 8-month period ending in 2016. 

\section{Neutron Detector and Pulse Shape Discrimination}
\begin{figure}
\includegraphics[width=0.45\textwidth,clip=]{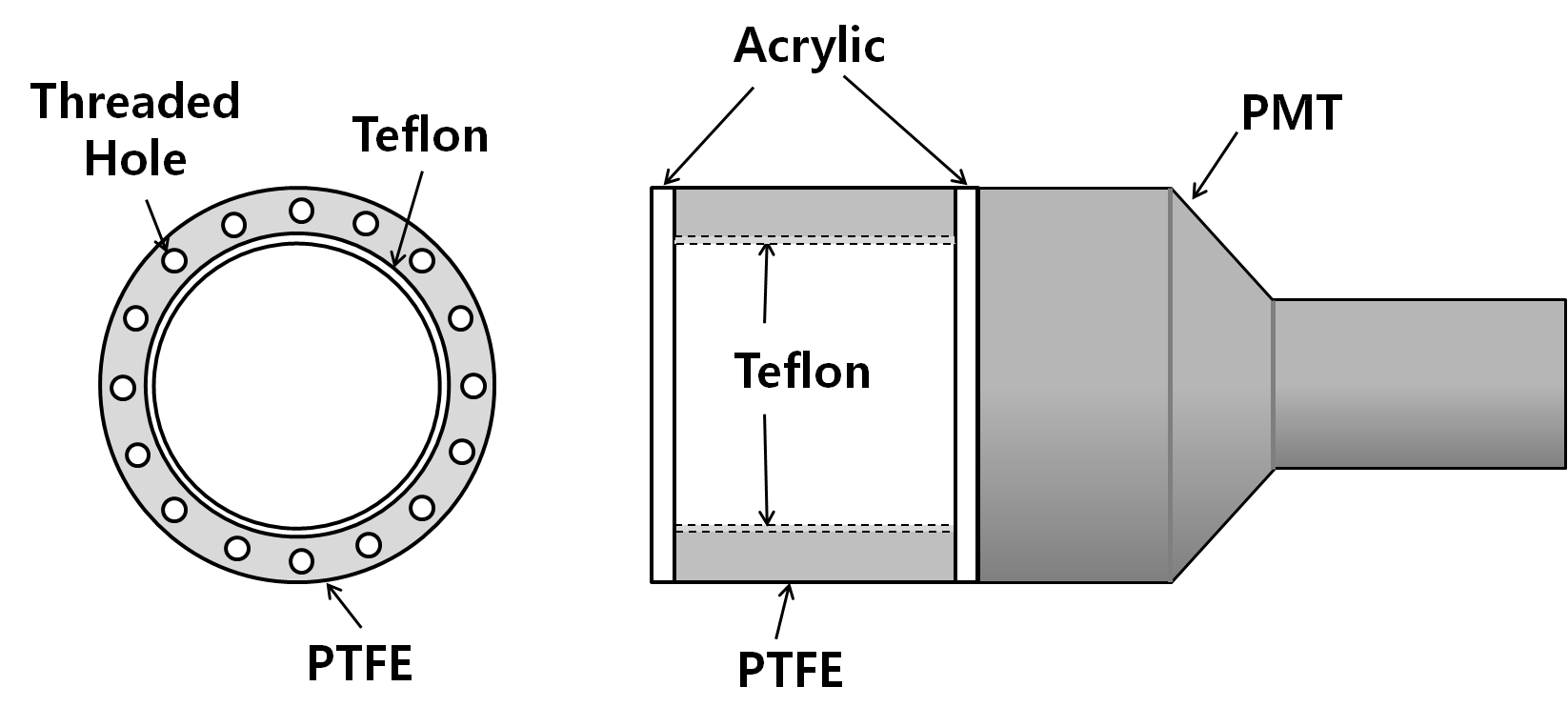}
\caption{The schematic drawing of the neutron detector. The cylindrical vessel with a Teflon
lining holding LS has an inner diameter of 10~cm and 
length of 10~cm. Each end of the cylinder is an $\sim 1$~cm thick transparent
acrylic plate. The vessel holds 0.78~liters of LS. A Hamamatsu R877-100 5-inch 
PMT is attached to the one end of the vessel.}\label{fig:neutrondetector}
\end{figure}

The neutron detector used in this work is a small volume LS
detector. The neutrons identification against dominant $\gamma$ background is done 
using a pulse shape discrimination (PSD) technique. 
The schematic drawing of the detector is shown in Fig.~\ref{fig:neutrondetector}. 
The detector uses a commercially available di-isopropylnaphtalene (DIN, C$_{16}$H$_{20}$) 
based LS, the Ultima Gold F\textsuperscript{TM} (PerkinElmer Inc.), for its high light 
yield with a good energy resolution and a good PSD performance. No diluent is used in LS. 
The PSD performance of the Ultima Gold F-based LS mixture has been studied in
Ref.~\cite{Kim:2015pba}.
The LS container vessel is made of a 10~cm long PTFE cylindrical pipe with an inner
diameter of 10~cm and a transparent cast acrylics plate on each end. The inner surface of 
the PTFE cylinder is lined with a Teflon sheet for an enhanced optical photon 
reflectivity. The vessel holds 0.78~liters of LS.
A Hamamatsu R877-100 5-inch photomultiplier tube (PMT) is attached to one of the 
acrylic plates. The whole 
detector assembly is then wrapped by black sheet to shield from external
optical photons. 
The signal from the PMT is pulse height triggered and digitized by a 12-bit FADC with a 500~MS/s
sampling rate.
A 400~ns delay is applied for the pedestal calculation and the time window for the charge 
integration is set to be 1~$\mu$s.
The PMT high voltage is set to yield $\sim 110$~pC/MeV.

The PSD technique uses the difference in the shape of the signal pulse due to the
different fluorescence characteristics of LS to different types of particles. The organic 
scintillators produce prompt and delayed fluorescence which decay times are 
on the order of nanoseconds for the prompt and hundreds of nanoseconds for the 
delayed.
The majority of light is produced by the prompt decay but the amount of the delayed 
decay varies depending on the type of particles causing the excitation. 
The neutron produced protons have a short range and generate high concentration
of triplet states that decay by delayed fluorescence, whereas electrons scattered
by $\gamma$'s have a longer range than protons and more likely to produce singlet
states which decay by prompt fluorescence. Here the amount of delayed fluorescence
is used to distinguish between the photons and neutrons.

\begin{figure}
\includegraphics[width=0.5\textwidth,clip=]{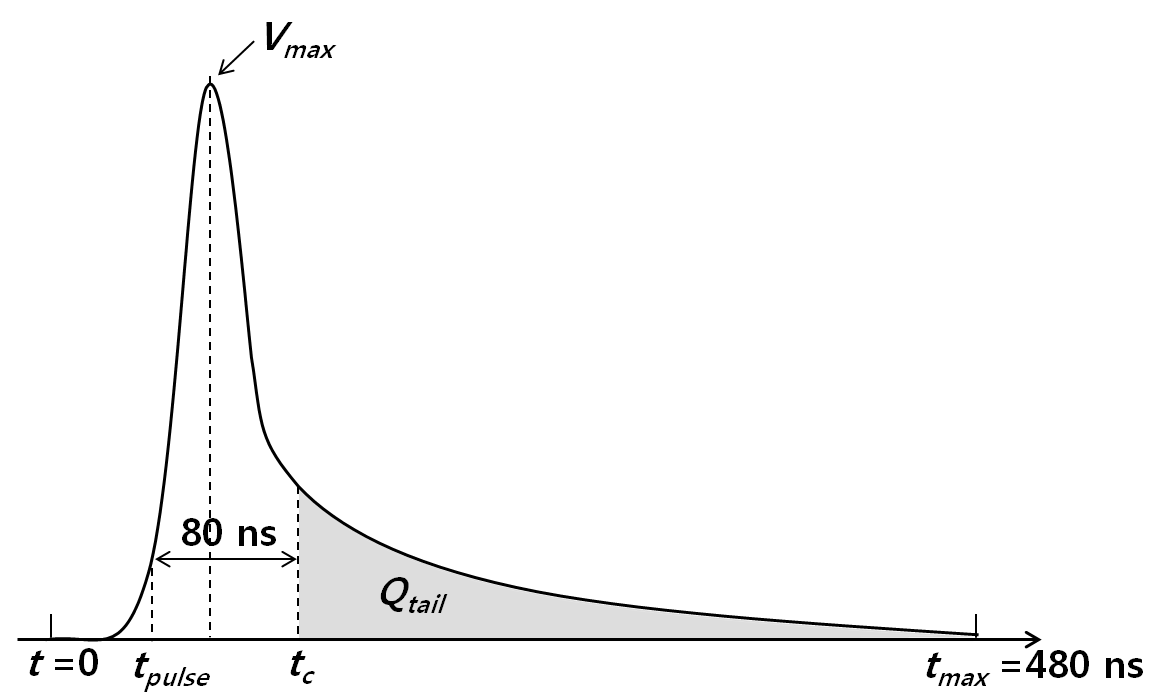}
\caption{An illustration of a pulse shape and the parameters used in calculation of
$Q_{tail}$. The shaded area is $Q_{tail}$ and the total area under the curve is
$Q_{total}$.}
\label{fig:qtail}
\end{figure}

The parameters used in the PSD method used here is illustrated in 
Fig.~\ref{fig:qtail}.
The tail charge $Q_{tail}$ is defined as 
\begin{equation}
  Q_{tail} = c \int_{t_c}^{t_{max}} V(t)\,dt,
\label{eq:qtail}
\end{equation}
where $V(t)$ is the height of the signal in voltage at time $t$, $c$ is the voltage-to-current
conversion constant, $t_{max}$ is the signal gate duration time (480~ns), and $t_c$ is the time 
defining the start of the tail section of the pulse, which is optimized for the neutron-$\gamma$ 
separation power. 
From the source calibrations described below,
$t_c$ is determined to be 80~ns after the pulse threshold time, $t_{pulse}$, that is defined by
\begin{equation}
  t_{pulse} = \sqrt{\ln{2}}\cdot w+{\bar t},
\end{equation}
where ${\bar t}$ and $w$ are parameters to be determined by the pulse shape of
each signal. The rising part of the pulse is fit with 
\begin{equation}
  V(t) = V_{max}\,\exp\left[-\left(\frac{t-\bar{t}}{w}\right)^2\right]~~(t<t_{V=V_{max}}),
\end{equation}
where $V_{max}$ is the maximum voltage in a pulse and $t_{V=V_{max}}$ is the time where the
maximum pulse height occurs.
The total charge $Q_{total}$ is obtained by integrating the pulse over the time 
range of $\left[0,~480\right]$~ns. 
The ratio of the charge in the tail to the total, $Q_{tail}/Q_{total}$,
is used as the PSD discriminant parameter.

\begin{figure}
\includegraphics[width=0.45\textwidth,clip=]{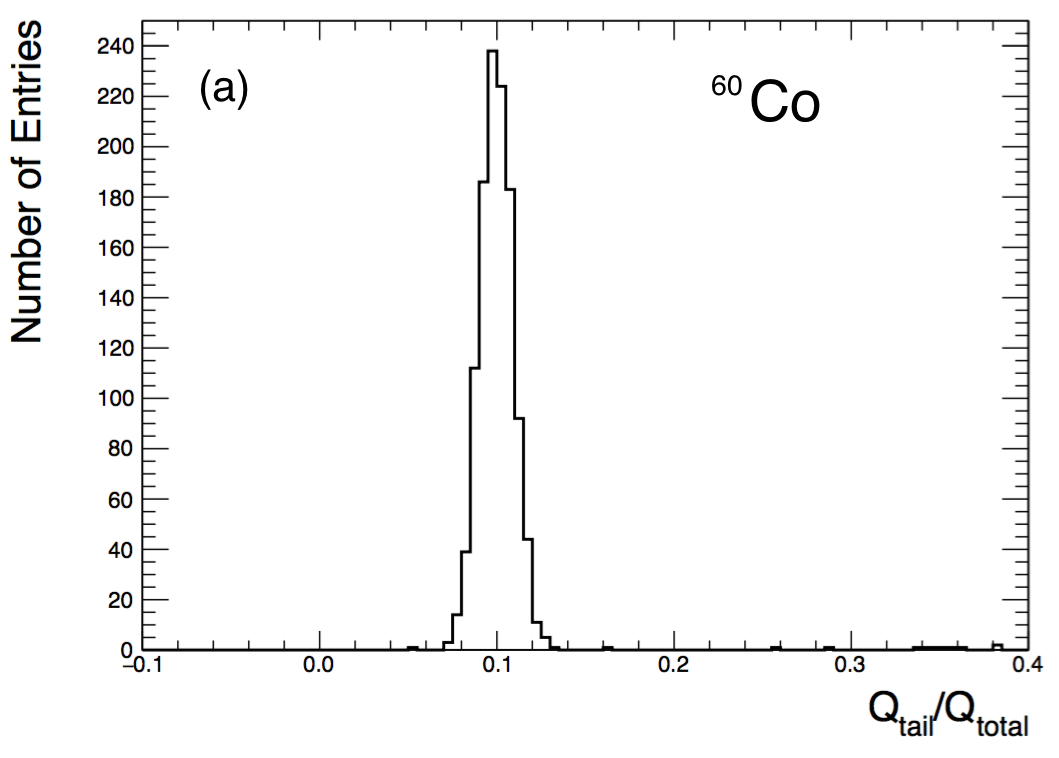}
\includegraphics[width=0.45\textwidth,clip=]{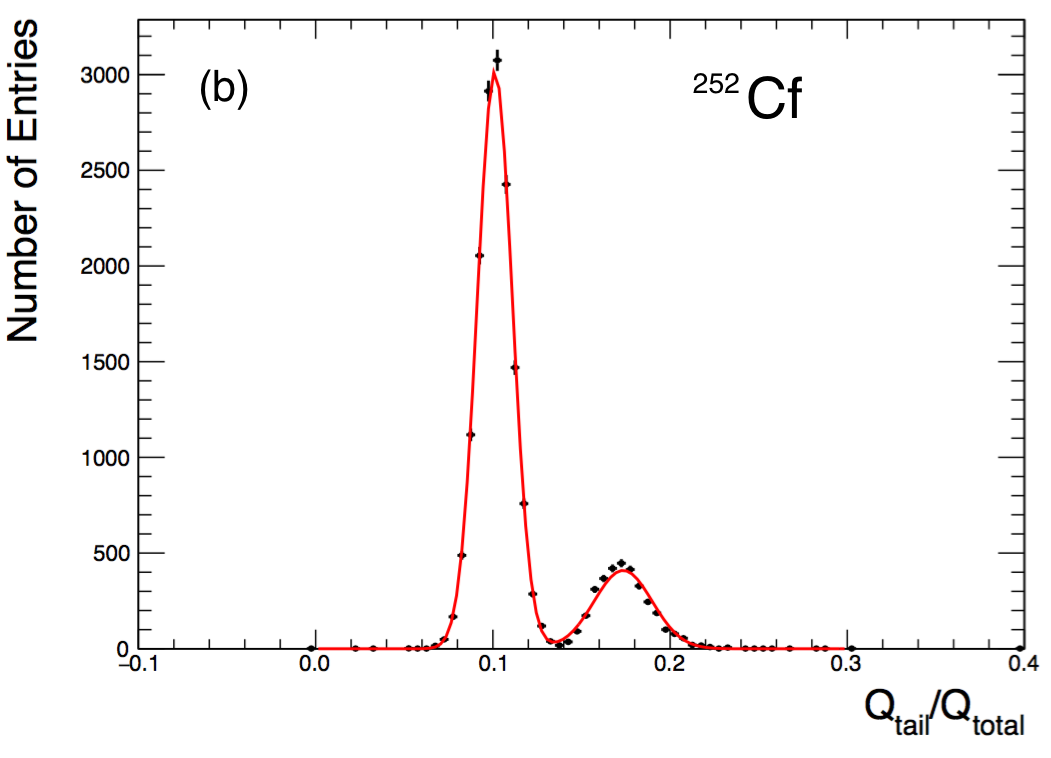}
\caption{The $Q_{tail}/Q_{total}$ distributions of (a) $^{60}$Co and 
(b) $^{252}$Cf. The red curve on (b) shows a fit with two Gaussian
functions. The figure of merit $S_p$ defined in Eq.~\protect\ref{eq:fom} is 
3.8 for the $^{252}$Cf distribution.}
\label{fig:sourcepsd}
\end{figure}

\begin{figure}
\includegraphics[width=0.45\textwidth,clip=]{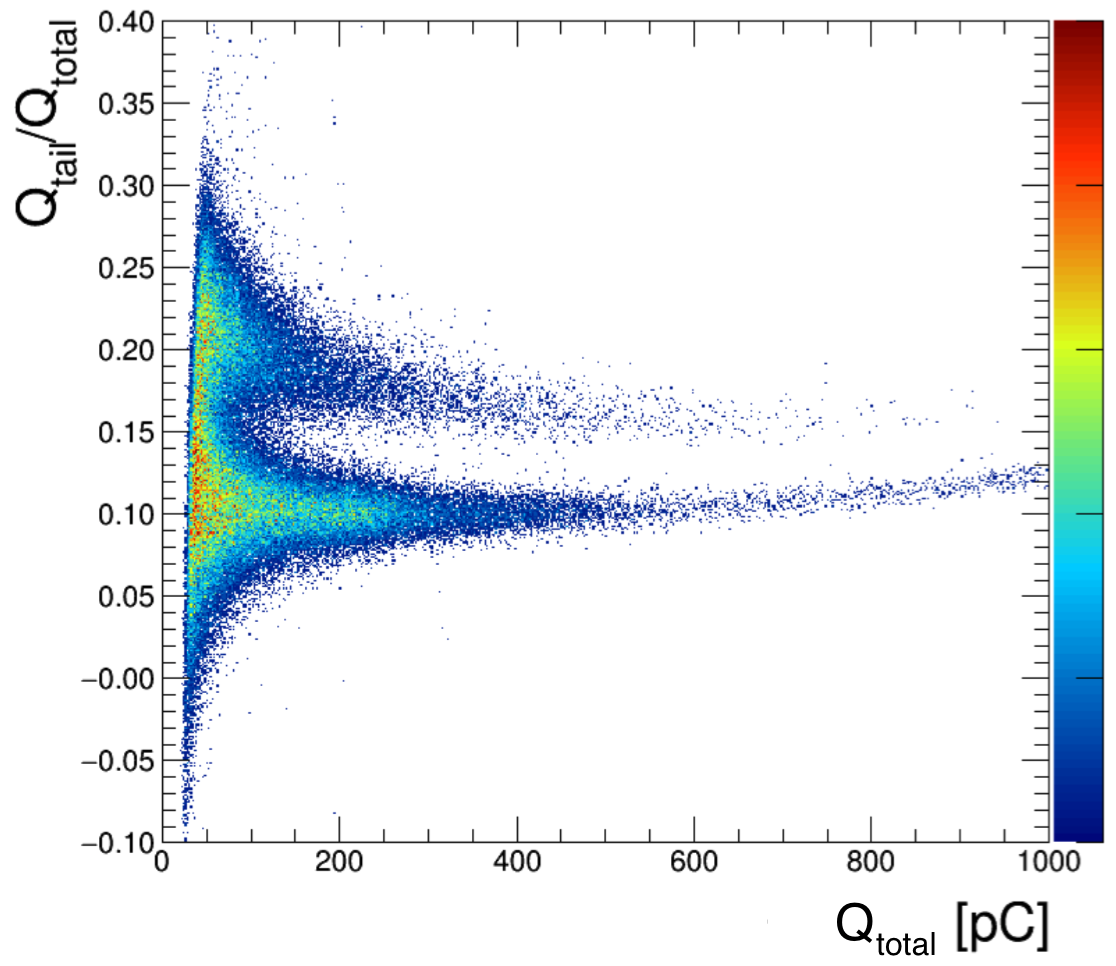}
\caption{(Color online) 2D scatter plot of $Q_{tail}/Q_{total}$ vs $Q_{tail}$ of the $^{252}$Cf   
source. A clear gap at $Q_{tail}/Q_{total}\sim 0.13$ in the event distribution is seen in 
the region of $Q_{total}$ over 100~pC. The upper distribution is neutrons and the lower one 
is $\gamma$'s. $Q_{total}$ dependence is clearly seen for $\gamma$'s.
}\label{fig:sourcepsd2}
\end{figure}

To test the performance of the detector, $^{60}$Co and $^{252}$Cf radioactive sources are used.
The $^{60}$Co source emits two $\gamma$'s with 1.17~MeV and 1.33~MeV. 
The $^{252}$Cf source emits an average of 3.7 neutrons per spontaneous fission with an
energy range of $\left[0, 13\right]$~MeV and the mean energy of $\sim 2$~MeV. 
Figure~\ref{fig:sourcepsd} shows the $Q_{tail}/Q_{total}$ distributions of data taken with 
$^{60}$Co and $^{252}$Cf sources. It shows a single peak at $Q_{tail}/Q_{total}=0.10$ 
for $^{60}$Co and two peaks at $0.10$ and $0.17$ for $^{252}$Cf
indicating $\gamma$'s and neutrons from the $^{252}$Cf source. 
The figure of merit for the power of separation is defined as
\begin{equation}
  S_p = \frac{\left|m_1-m_2\right|}{\sigma_1+\sigma_2},
\label{eq:fom}
\end{equation}
where $m_1$ and $m_2$ are the mean values and $\sigma_1$ 
and $\sigma_2$ are the widths of the two Gaussian 
functions fit to the $Q_{tail}/Q_{total}$
distribution. The parameter $t_c$ is obtained from maximizing $S_p$. 
Figure~\ref{fig:sourcepsd2} the scatter plot of $Q_{tail}/Q_{total}$ vs $Q_{total}$. A clear
separation is seen over $Q_{total}=100$~pC.

\section{Neutron Rate Measurements}
The neutron rates are measured at two sites; the tendon gallery of the Hanbit nuclear 
power plant reactor unit 5 complex 
in Yeonggwang where the NEOS detector is placed and a site known as the KT1 laboratory in Daejeon,
both in Korea. 
The tendon gallery of the reactor complex is located 10~m below the surface and has 
an overburden of about 20~m water equivalent.
The KT1 laboratory has a negligible amount of overburden as is the
case for many research reactor facilities.

\begin{figure}
\includegraphics[width=0.45\textwidth,clip=]{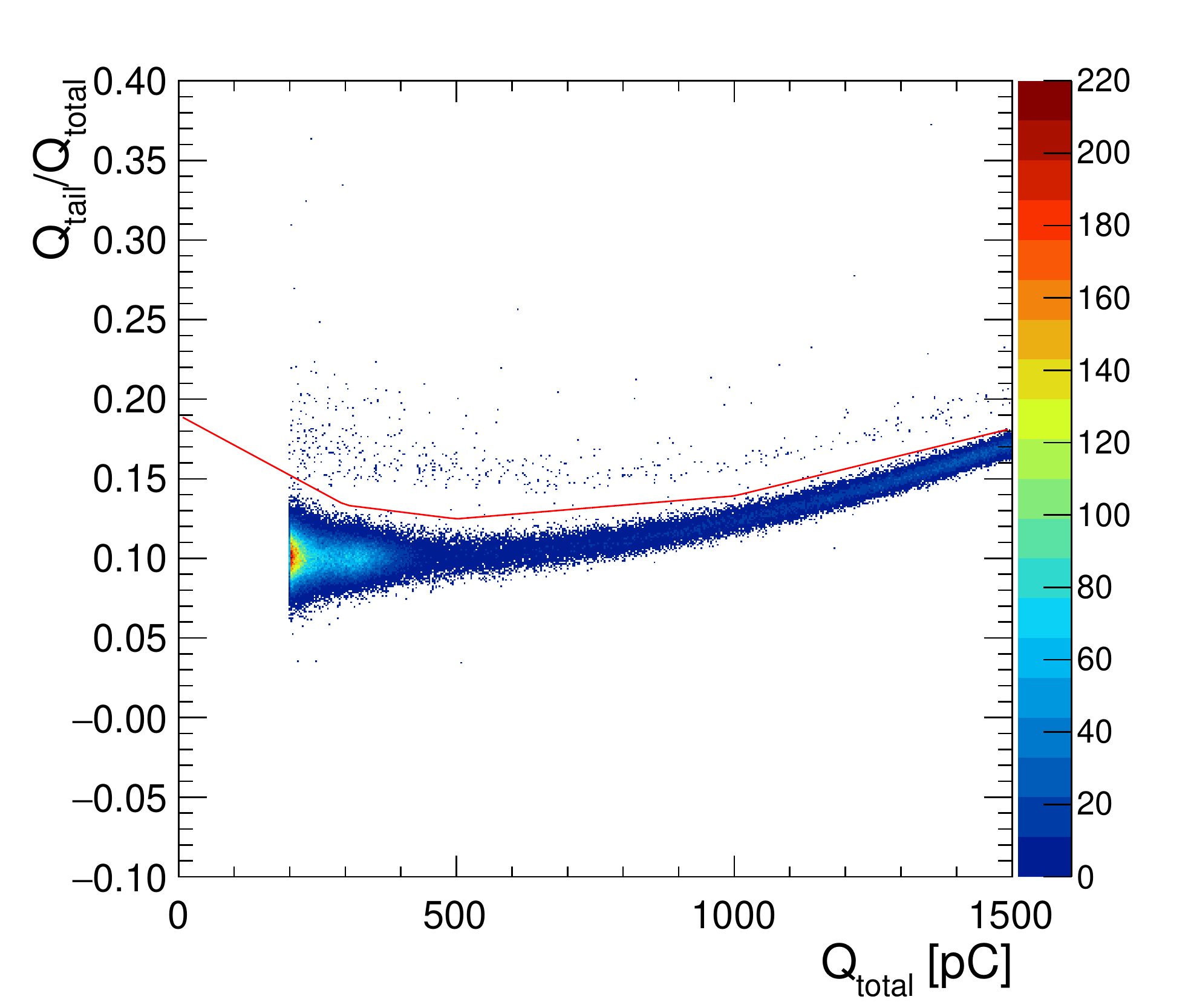}
\caption{(Color online) 2D scatter plot of $Q_{tail}/Q_{total}$ vs $Q_{total}$ 
at KT1 laboratory for a data taken for $\sim 2$ hours. 
The events above the red line and $200<Q_{tail}/Q_{total}<1\,500$~pC 
are counted as neutron events.}\label{fig:kt1psd}
\end{figure}

\begin{figure}
\includegraphics[width=0.45\textwidth,clip=]{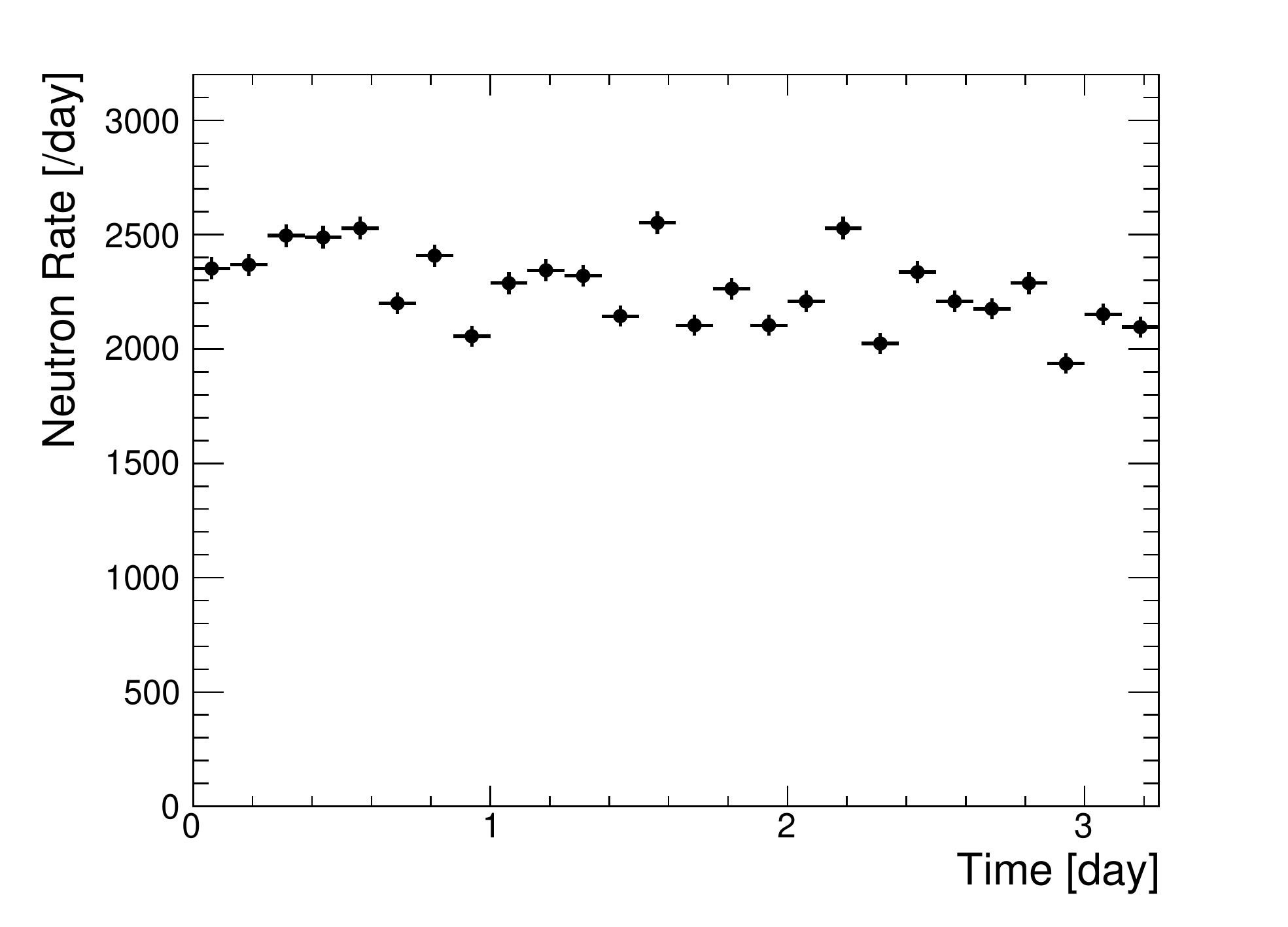}
\caption{The neutron rates measured at KT1 laboratory. Each point represents
about a three hour long data.}\label{fig:kt1_rates}
\end{figure}

The data is taken with the detector for three days at the KT1 laboratory with a 5 cm thick lead 
shielding to reduce the external $\gamma$'s. 
Figure~\ref{fig:kt1psd} shows the scatter plot of $Q_{tail}/Q_{total}$ vs $Q_{total}$
for a set of data taken for about two hours at KT1 laboratory.
A separation between neutrons and $\gamma$'s can be seen along $Q_{tail}/Q_{total}$. 
The red line shown in
the plot separates the neutrons from $\gamma$'s, which is determined for 
each data set. The events in $200 < Q_{tail}/Q_{total} < 1\,500$~pC are counted.
The measured neutron rate is about $2\,300$ per day as shown in Fig.~\ref{fig:kt1_rates}. 

\begin{figure}
\includegraphics[width=0.45\textwidth,clip=]{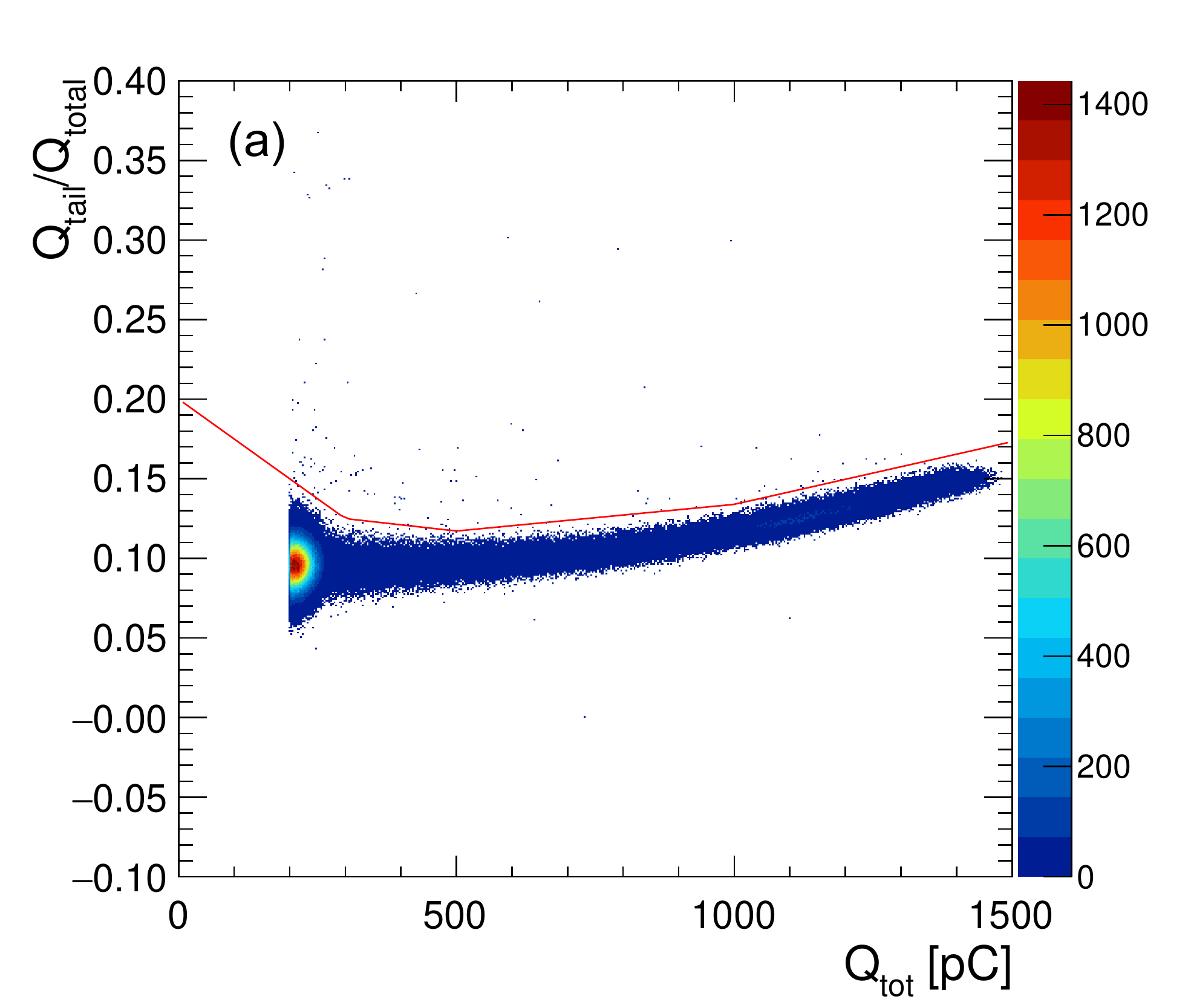}
\includegraphics[width=0.45\textwidth,clip=]{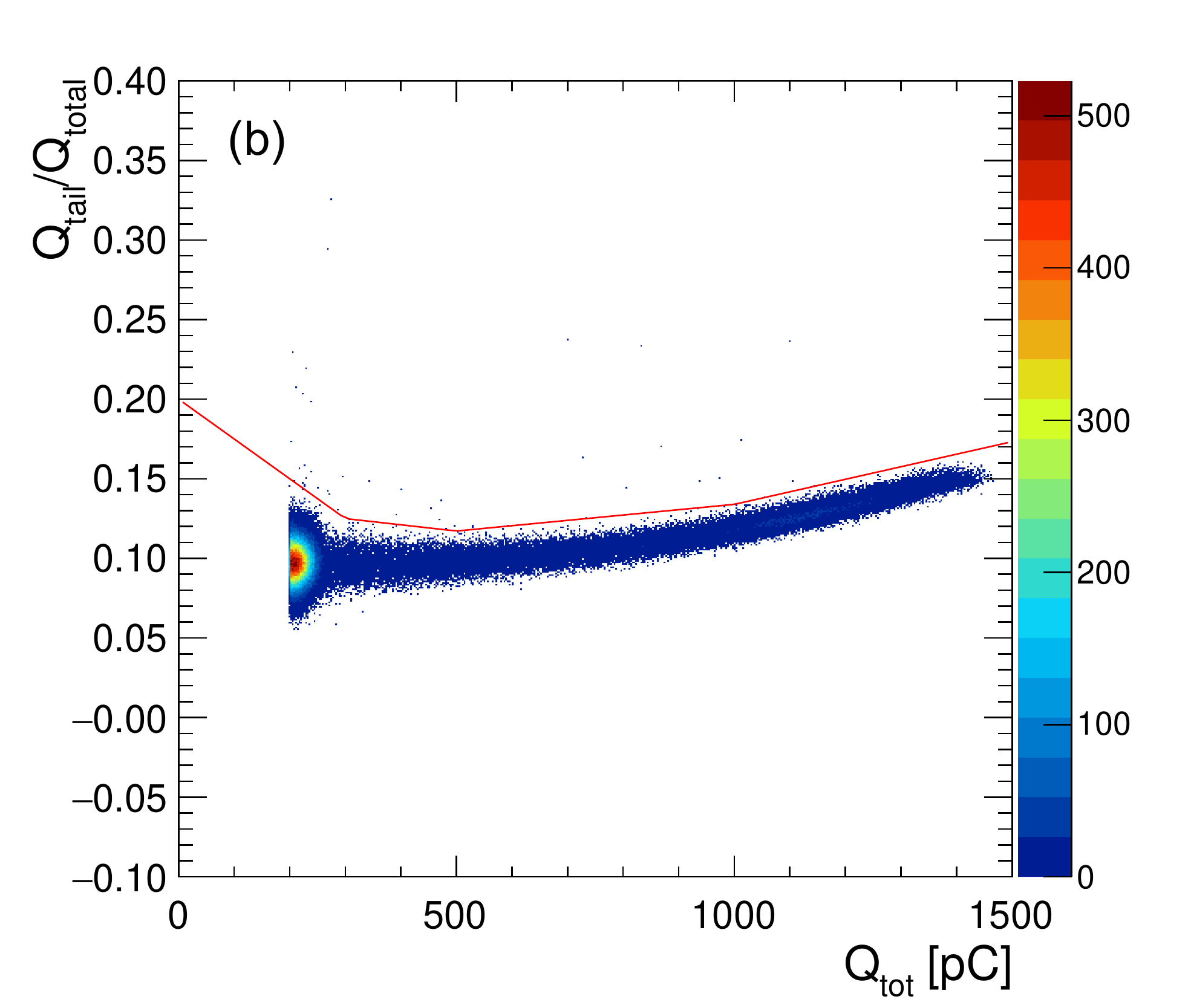}
\caption{(Color online) 2D scatter plots of 
$Q_{tail}/Q_{total}$ vs $Q_{total}$ 
at the tendon gallery of reactor building complex during (a) reactor-off and (b) reactor-on periods. 
The duration of the data taking for the reactor-off plot is about three times longer 
than that of the reactor-on plot shown here.
The events above the red line are counted as neutron events.}
\label{fig:tendonpsd}
\end{figure}
\begin{figure}
\includegraphics[width=0.45\textwidth,clip=]{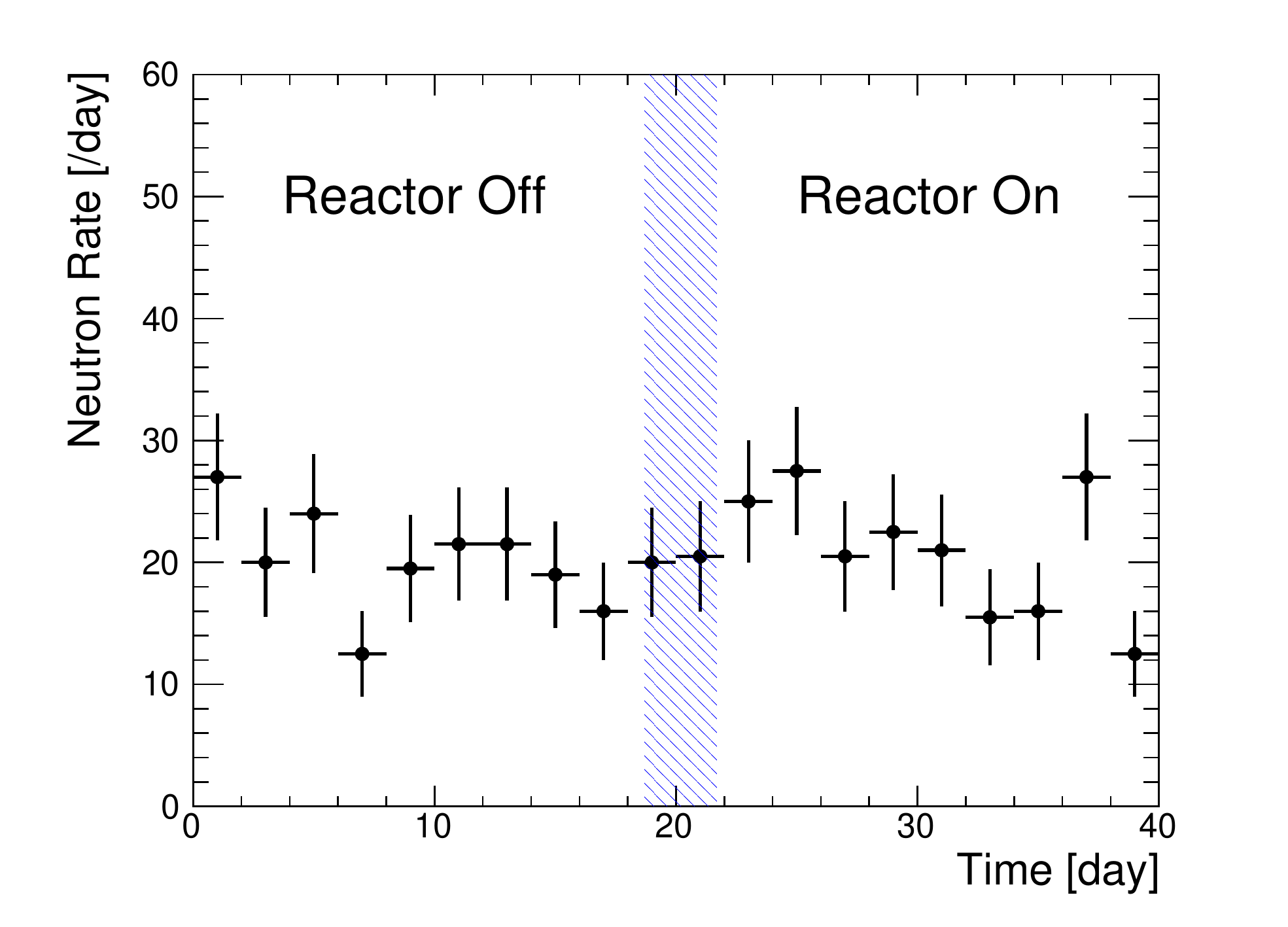}
\caption{The neutron rates at the tendon gallery of Hanbit nuclear 
power plant reactor complex during
the reactor-on and off periods. The shaded region is the reactor ramp on period.}
\label{fig:tendon_neutron_rates}
\end{figure}

The measurements is made at the tendon gallery of Hanbit nuclear power plant 
complex for a period of about 40 days, which includes periods of 
reactor-off and on as well as ramping up. Figure~\ref{fig:tendonpsd} shows the 
scatter plots of $Q_{tail}/Q_{total}$ vs $Q_{total}$ during the reactor-on and 
off periods. 
The fast neutron rates are measured to be $20.0\pm 1.82$ and $21.8\pm 1.58$ 
per day for the reactor-off and on periods, respectively.
No significant difference is observed between reactor-on and off periods indicating 
that there is enough shielding against neutrons originating from the reactor at the 
tendon gallery.
Figure~\ref{fig:tendon_neutron_rates} shows the fast neutron rates 
during the reactor-on, off, and ramping up periods.
The results are summarized in Table~\ref{tb:result}.
The variations in the neutron selection cut in $Q_{tail}/Q_{total}$ is accounted for
as a systematic uncertainty.
\begin{table}
\caption{Results of neutron rate measurements at KT1 laboratory and tendon gallery with
reactor off and on.}
\begin{tabular}{cccc} \hline\hline
& Live Time [days]   & Neutron Event Rate [/day] \\\hline
KT1 Laboratory       & $3.36$ & $2\,261\pm   26({\rm stat.})\pm 116({\rm syst.})$\\
Tendon Gallery (off) & $18.7$ & $  20.0\pm 1.03({\rm stat.})\pm 1.50({\rm syst.})$\\
Tendon Gallery (on)  & $17.5$ &$   21.8\pm 1.12({\rm stat.}) \pm 1.12({\rm syst.})$\\\hline
\end{tabular}
\label{tb:result}
\end{table}


\section{Conclusion}
The fast neutron rate is measured with a small LS detector using the PSD technique. The 
fast neutron rate at the tendon gallery of Hanbit nuclear power plant where the NEOS
experiment is performed is measured to be $\sim 1/100$ of that of the
overground facility with a negligible overburden. It is also found that the reactor operation 
does not affect the fast neutron rate at the NEOS experiment site.


\begin{acknowledgments}
This work is supported by IBS-R016-D1 and 2012M2B2A6029111 from National Research Foundation
(NRF). We appreciate the assistance from the Korea Hydro and Nuclear Power (KHNP) company, 
especially acknowledge the Safety and Engineering Support Team of Hanbit-3 Nuclear Power Plant.
\end{acknowledgments}

\end{document}